\documentclass{appolb}
\usepackage{graphicx,amsmath}
\usepackage{wrapfig,url}
\usepackage{lineno}

\begin{document}
\title{Direct measurement of photons from the electron-hadron bremsstrahlung at the EIC}

\author{L. Adamczyk, Y. Ali\footnote{Now at COMSATS University, Islamabad, Pakistan}, A. B. Kowalewska\footnote{Now at Jagiellonian University, Kraków, Poland}, B. Pawlik,\\ K. Piotrzkowski\thanks{Corresponding author: piotrzkowski@agh.edu.pl}, M. Przybycien
\address{AGH University of Krakow, Faculty of Physics and Applied Computer Science, al. Mickiewicza 30, 30--059 Kraków, Poland}
\\[3mm]
J. J. Chwastowski
\address{Institute of Nuclear Physics PAN, ul. Radzikowskiego 152,
31--342 Krak\'ow, Poland}
}


\maketitle

\begin{abstract}
Direct detection of bremsstrahlung photons, in principle, offers the most straightforward and most robust method of luminosity determination at the EIC, but requires an extraordinary performance of the photon detector. In this paper, we first discuss the extreme working conditions for such detectors at the EIC and the resulting technology choices. 
Then, we report the initial results of Monte Carlo simulations, using {\sc Geant4}, of the proposed sampling calorimeter, which is made of a copper absorber with embedded quartz fibres read out by silicon photomultipliers. Finally, the tentative requirements for appropriate readout electronics are formulated.
\end{abstract}



\section{Introduction}
\label{sec:intro}

Precise measurements of the electron-hadron cross sections are the cornerstone of scientific programme at the future Electron-Ion Collider (EIC), hence high demands on the EIC luminosity measurements – at least a 1\% accuracy is required for the absolute luminosity determination and only a $10^{-4}$ uncertainty is expected for the relative (bunch-to-bunch) luminosity measurements, relevant for precise spin asymmetry studies at the EIC~\cite{AbdulKhalek:2021gbh}. It was demonstrated at HERA – the first electron-hadron collider -- that a direct bremsstrahlung measurement can be used successfully to precisely determine the absolute luminosity of high-energy $ep$ collisions~\cite{ZEUSLuminosityMonitorGroup:1992hmm, Piotrzkowski:1993zv, ZEUSLuminosityGroup:2001eva}. As discussed below, this technique can also be applied at the EIC; however, it poses a significant challenge due to the enormous bremsstrahlung rates. At the same time, it offers an exceptional statistical power for ultra-precision measurements. Typically, more than a billion high-energy bremsstrahlung photons can be directly measured at the EIC every second.
\begin{figure}[b]
\centering
\includegraphics[width=.28\textwidth]{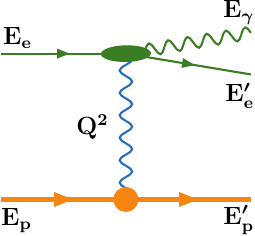}\hspace{11mm}
\includegraphics[width=.56\textwidth]{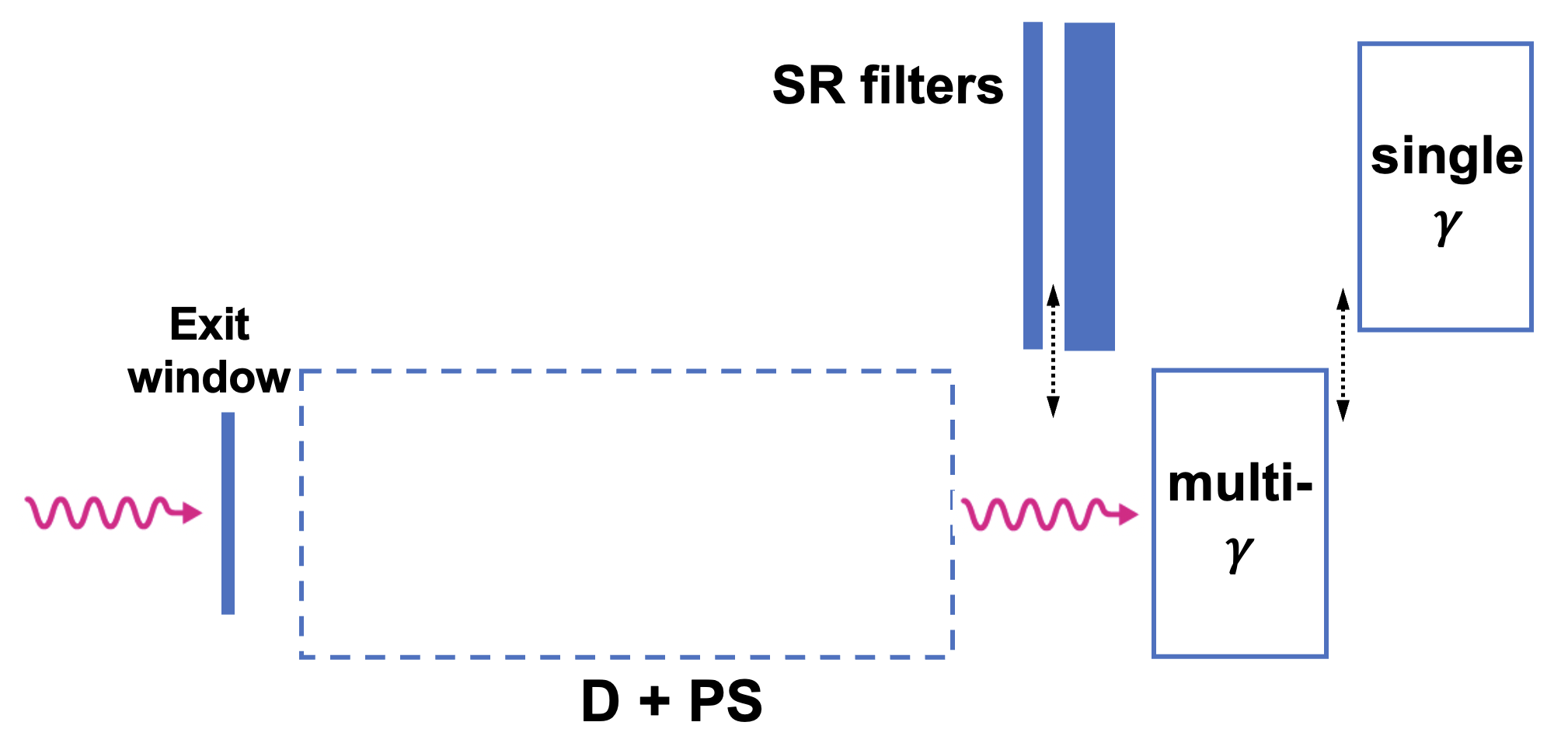}
\caption{(left) The bremsstrahlung diagram in the leading order, and (right) a simple sketch of the layout of the bremsstrahlung detectors at the EIC with two movable calorimeters for direct photon detection and two movable synchrotron radiation filters (SR), where the dashed rectangle represents the $e^+e^-$ pair spectrometer, PS, for the detection of converted photons, which includes two dipole magnets D.
\label{fig:i}}
\end{figure}

In Fig.~\ref{fig:i}, a generic diagram for the bremsstrahlung process is shown, as well as
a simple sketch of the layout of bremsstrahlung photon detectors at the EIC. 
Following the ZEUS design at HERA~II~\cite{ZEUS:2013emt}, an indirect measurement of bremsstrahlung will also be performed, utilising the $e^+e^-$ pairs from the photon conversion. Bremsstrahlung photons will leave the EIC primary vacuum vessel via an exit window at about 20~m from the interaction point (IP)~\cite{AbdulKhalek:2021gbh}, then the electrons and positrons from photon conversion will be swept away by the field of dedicated dipole magnets, so with a very good approximation only unconverted bremsstrahlung photons reach the calorimeter behind, which can be protected by movable filters against direct synchrotron radiation (SR), when necessary. Finally, it has been proposed~\cite{Piotrzkowski:2021cwj} to use two small movable calorimeters for direct measurements that can be placed alternately in the working position. 

\begin{figure}[t]
\centering
\includegraphics[width=.48\textwidth]{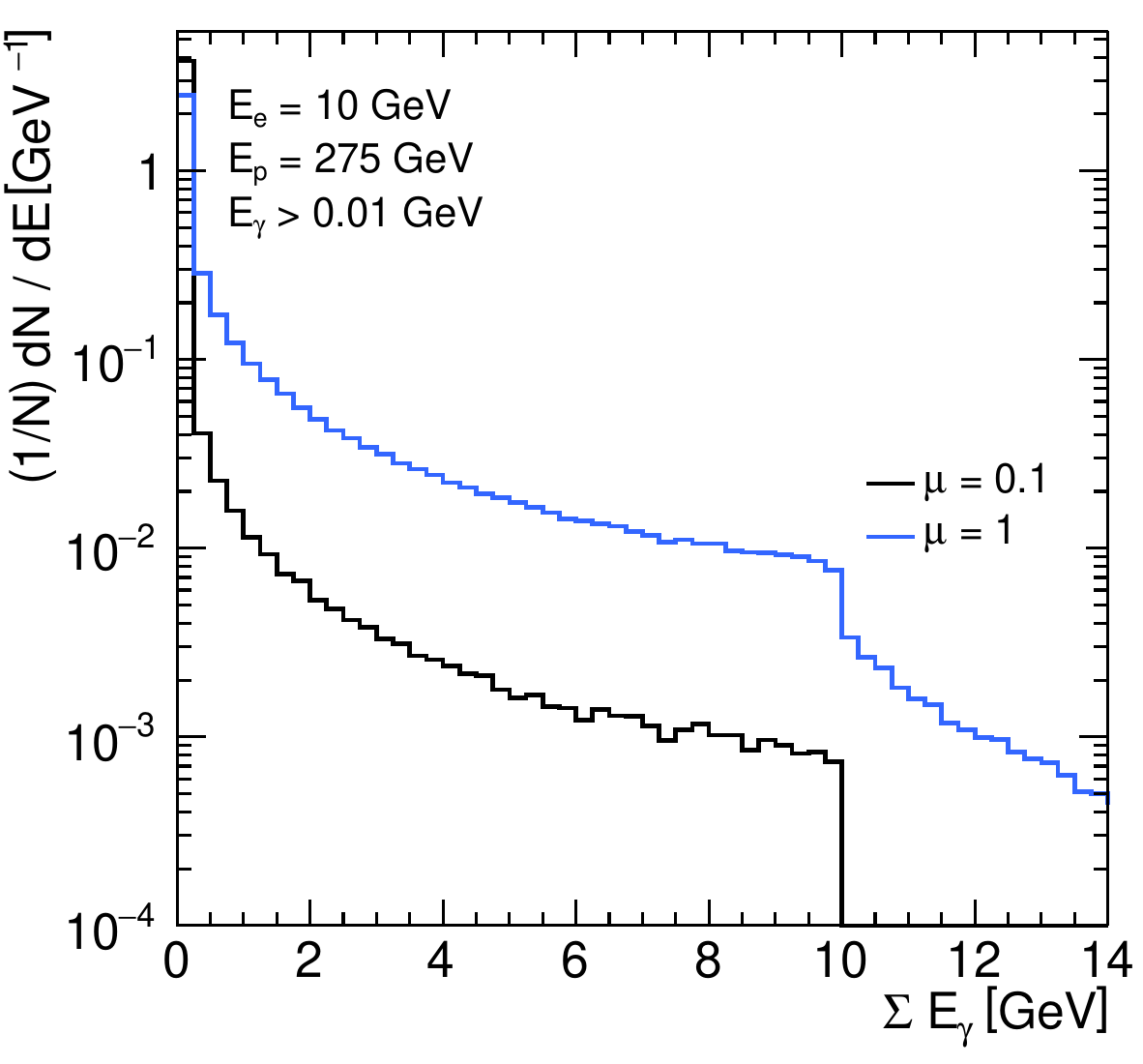}
\hspace{2mm}
\includegraphics[width=.48\textwidth]{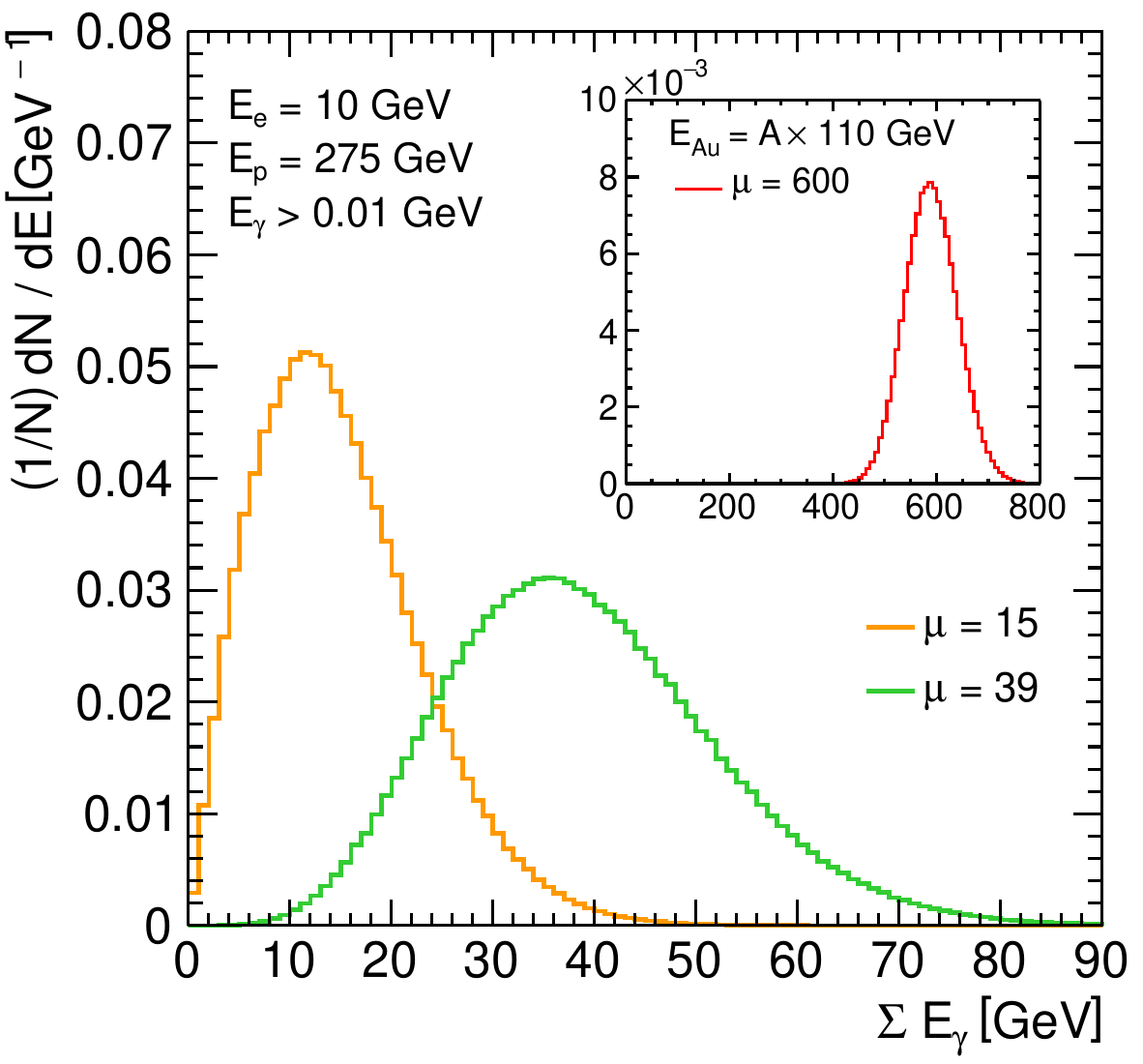}
\caption{Distribution of the total energy of bremsstrahlung photons per bunch crossing at low (left) and high (right) EIC luminosity, assuming the photon threshold energy of 10 MeV. The nominal $ep$ luminosity of $10^{34}\;\rm cm^{-2}s^{-1}$ corresponds to an average number of such photons $\mu=39$. The distribution in the insert corresponds to the $e$Au luminosity of $5\cdot10^{33}$ cm$^{-2}$s$^{-1}$~per nucleon.
\label{fig:ii}}
\end{figure}

The calorimeter ``multi-$\gamma$" (MGC) will be optimised for measurements at high luminosity ${\cal L}$, and the calorimeter ``single-$\gamma$" (SGC) will be used only at lower luminosities, below $10^{33}\;\rm cm^{-2}s^{-1}$ in the case of $ep$ collisions, during the first years of EIC running, as well as during dedicated calibration runs. 
The geometrical acceptance of bremsstrahlung photons will be very high at the EIC, at least 99\%~\cite{AbdulKhalek:2021gbh}. 
As a result, the rate of bremsstrahlung photons hitting the direct detectors will be enormous, especially in the case of the electron collisions with heavy ions, as the bremsstrahlung cross-section scales with the nucleus charge squared.
Typically, the beam bunches will collide at the EIC every 10 ns, but a significant event pile-up will occur; see Fig. \ref{fig:ii}.
For example, for the nominal $ep$ luminosity of $10^{34}\;\rm cm^{-2}s^{-1}$, and the 10 GeV and 275~GeV electron and proton beams, respectively, on average 39 photons with energies above 10~MeV will hit a direct-photon detector every bunch collision, and about 600 such photons for the nominal $e$Au collisions! 

This means that at high luminosity, its measurement effectively corresponds to measuring an average energy deposited in the calorimeter. In contrast, the luminosity measurements at low $\cal L$ can be performed using simple photon counting, which has been very successfully employed at HERA~I. 
In this case, the energy cutoff of the photon spectra at the beam energy (see Fig. \ref{fig:ii}) can be used for a very precise data-driven calibration of the photon energy scale~\cite{Piotrzkowski:1993zv, ZEUSLuminosityGroup:2001eva, ZEUS:2013emt}, and consequently the SGC can provide precision reference measurements at low $\cal L$.
\section{Working conditions at the EIC}

The bremsstrahlung rate will be so high at the EIC that, with a very good approximation, one can assume that this will be the only relevant source of radiation damage for the direct-photon detectors. Direct synchrotron radiation will be stopped entirely by the thick exit window in the case of a 5~GeV electron beam, and only for the highest electron beam energy of 18 GeV, a graphite filter of about one radiation length could be required~\cite{PDR}. In the following initial studies for the nominal 10~GeV electron beam, we simulate the bremsstrahlung photons directly hitting the photon calorimeters.

The SGC should have the best possible energy resolution to allow for precise measurements of bremsstrahlung spectra and reference luminosity measurements~\cite{Piotrzkowski:2021cwj}. As one is also dealing with low-energy photons, well below 1 GeV, a homogeneous calorimeter made of scintillator is a natural SGC candidate. In the following, we show the results obtained using {\sc Geant4}~\cite{GEANT4:2002zbu,Allison:2016lfl} for the bremsstrahlung-induced doses in a 20 cm $\times$ 20 cm $\times$ 30 cm block of PbWO$_4$ crystal. The bremsstrahlung photons were generated using {\sc Bremge} (see appendix) and hit the calorimeter along the z-axis with the transverse impact point Gaussian-smeared, with $\sigma_x=\sigma_y=1$~cm, to effectively account for the photon angular distribution at the distant interaction point over time.
The choice of the lead tungstate is dictated by its relatively fast light output, which is necessary at the EIC. This crystal has been used in the construction of the electromagnetic calorimeter in the CMS experiments at CERN, and comprehensive studies of its radiation hardness have shown a significant suppression of light output for doses above 1 kGy~\cite{CMS:2009heh}.

\begin{figure}[b]
\centering
\includegraphics[width=.55\textwidth]{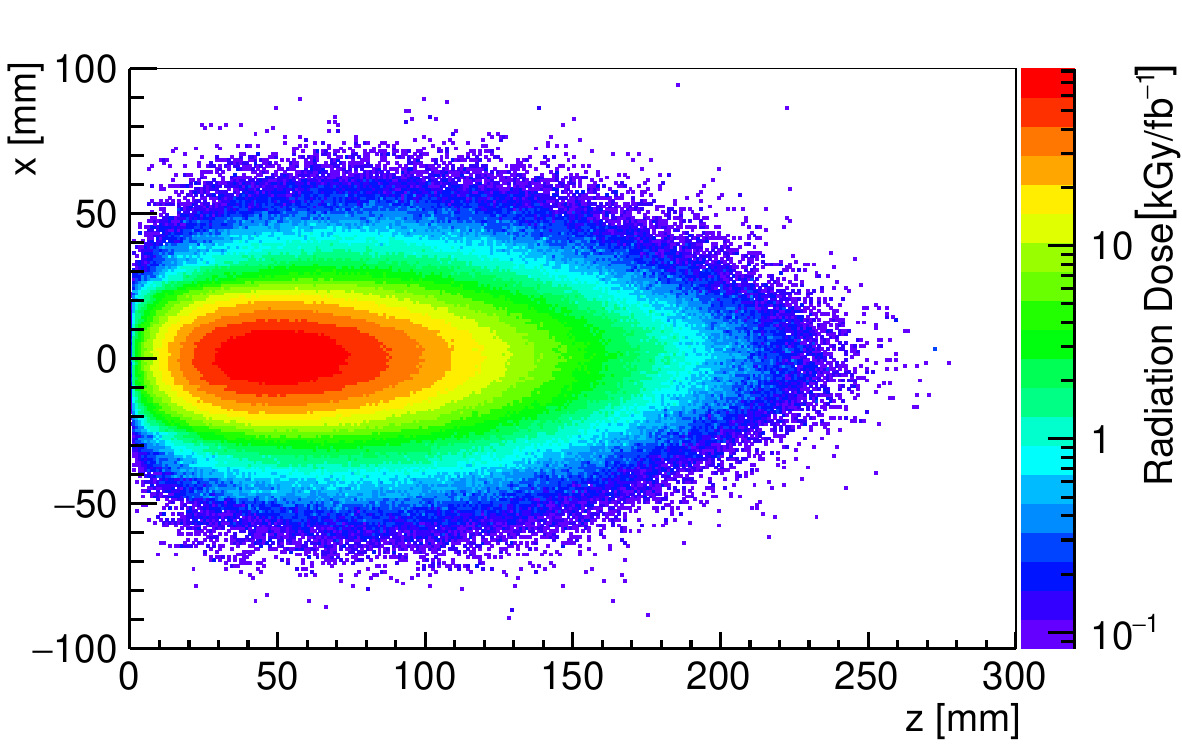}
\hspace*{10pt}\includegraphics[width=.39\textwidth]{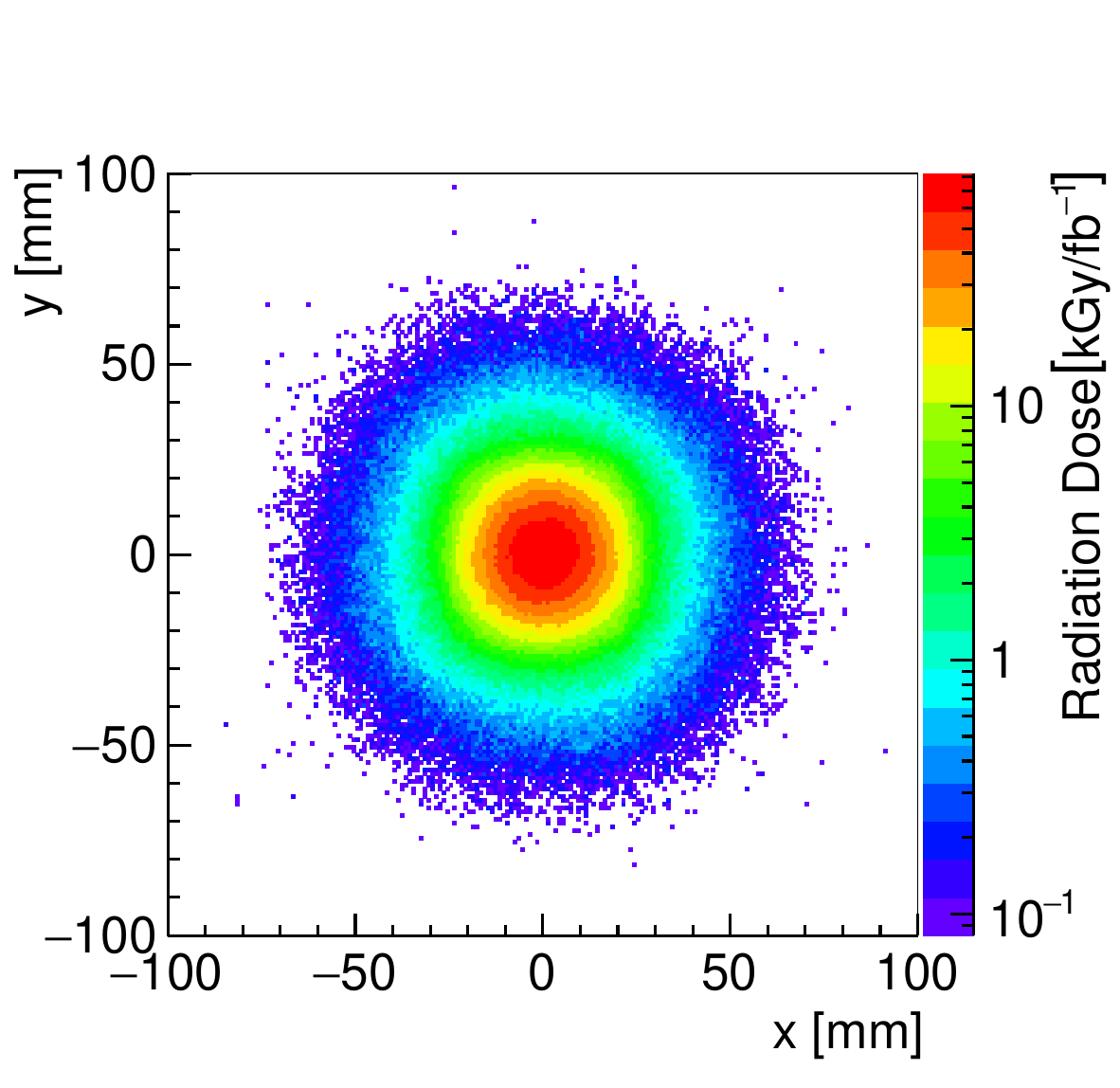}
\caption{(left) Distributions of the dose per 1 fb$^{-1}$ due to bremsstrahlung in a PbWO$_4$ calorimeter, assuming the 10 GeV and 275 GeV electron and proton beams, respectively -- the longitudinal dose distribution for $y=0$, and (right) the transverse distribution at the maximum of the longitudinal distribution.
\label{fig:iii}}
\end{figure}

According to the results shown in Fig. \ref{fig:iii}, such a crystal calorimeter could be used to measure the electron-proton bremsstrahlung for an integrated luminosity of at most 10~pb$^{-1}$, which corresponds to approximately 12 days of continuous operation at a low luminosity of $10^{32}\rm cm^{-2}s^{-1}$. In addition, since bunch collisions at the EIC usually occur every 10~ns, to avoid an overlap of subsequent scintillator light pulses, it is then necessary to run only with $\mu\ll1$. Therefore, such an SGC could only be used for short calibration runs at very low $\cal L$, and if a better radiation hardness is required and the SGC's operation at a higher luminosity, for example, a sampling calorimeter with fast plastic scintillators can be used instead.

\begin{figure}[t]
\centering
\includegraphics[width=.48\textwidth]{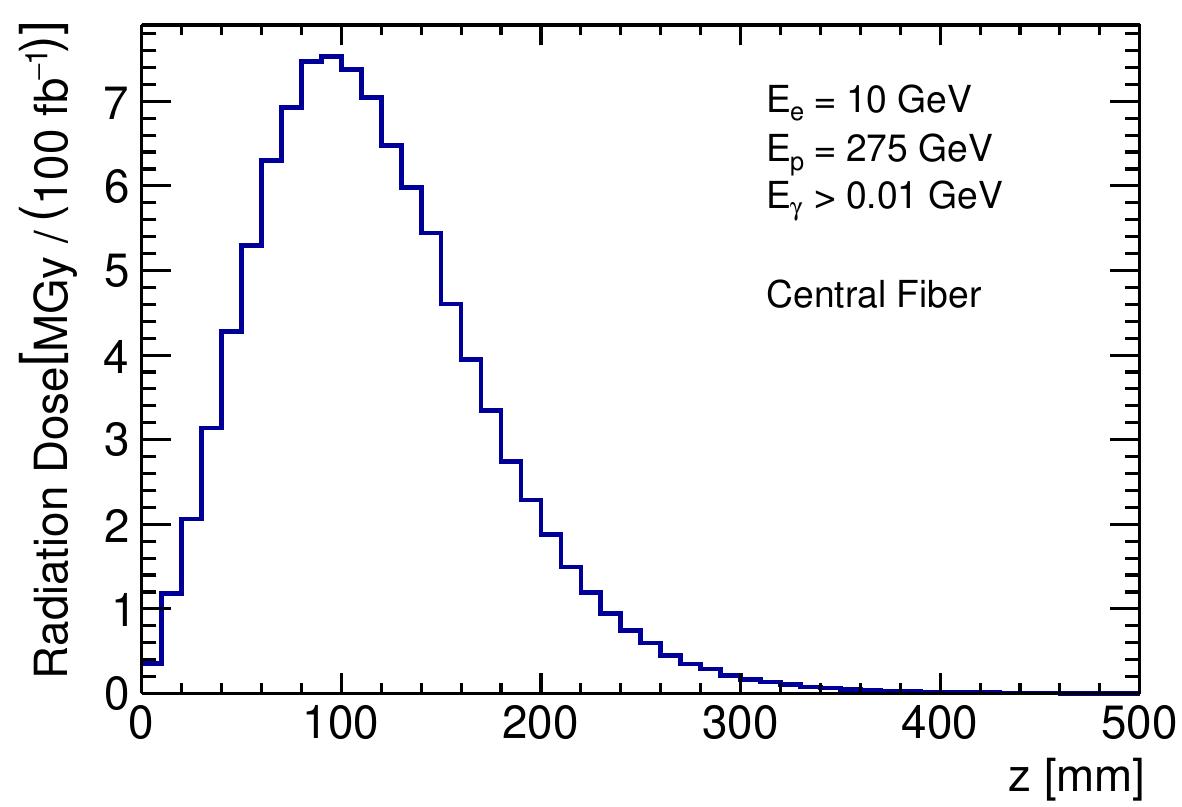}
\hspace{1mm}
\includegraphics[width=.48\textwidth]{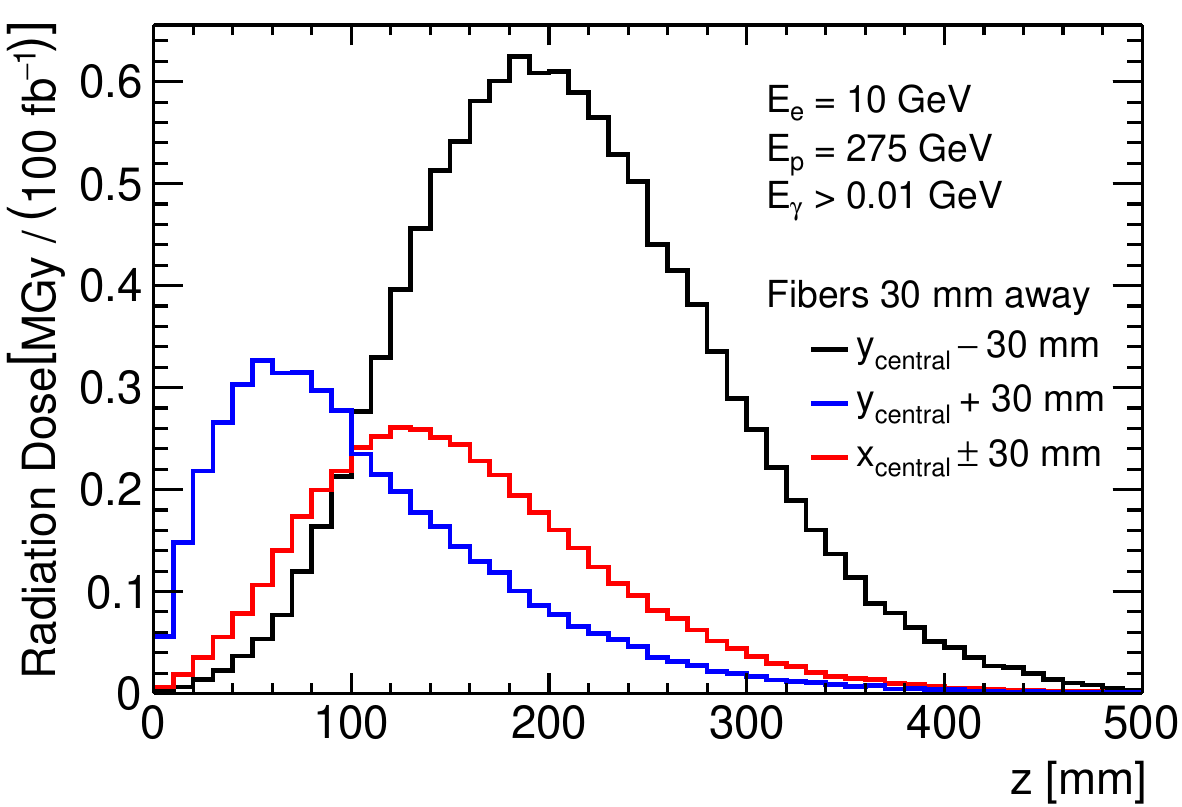}
\caption{(left) Distributions of the dose due to bremsstrahlung in central fibres of copper-quartz ``spaghetti" calorimeter, assuming a 10 GeV electron beam colliding with 275 GeV protons per 100 fb$^{-1}$, and (right) in fibres 30 mm away horizontally from the central fibre.
The calorimeter is tilted vertically by 5 degrees.\label{fig:iv}
}
\end{figure}

At nominal luminosity, a much more radiation-tolerant calorimeter technology must be used for the direct bremsstrahlung measurements at the EIC using the MGC. In the following, we consider a sampling calorimeter made of a copper absorber and longitudinally embedded quartz fibres (a {\it QFi} "spaghetti" calorimeter). The choice of copper (instead of tungsten, for example) is motivated by the need for a very efficient dissipation of the heat produced by bremsstrahlung absorption, at the rate of about 0.7 and 10~W for the nominal $ep$ and $e$Au collisions, respectively. In addition, as the electromagnetic cascades are then relatively wide, it somewhat diminishes the maximal doses in central fibres.

\begin{wrapfigure}{r}{0.35\textwidth}
\centering
\includegraphics[width=.34\textwidth]{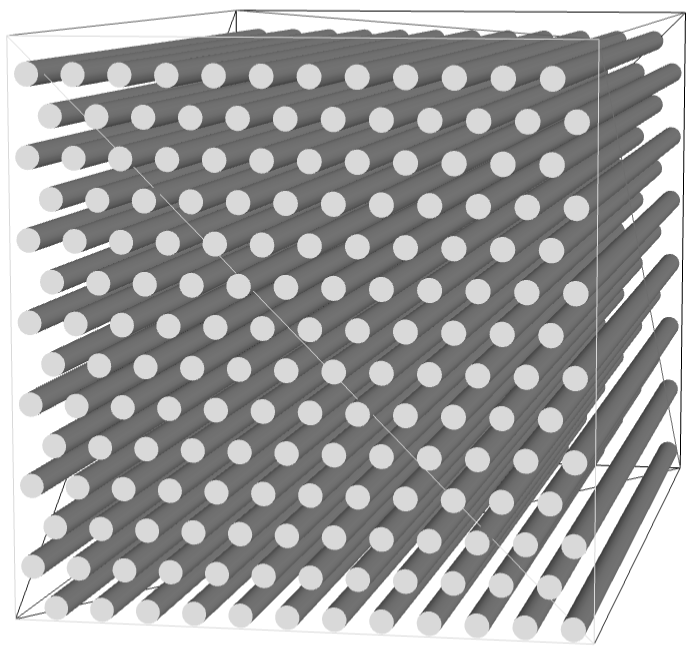}
\caption{Pattern of fibres.\label{fig:honey}
}
\end{wrapfigure}
The quartz fibres in the MGC are laterally distributed in a honeycomb pattern (see Fig.~\ref{fig:honey}) with a distance between the closest fibres of 2~mm. The fibres are made of a 1.2~mm diameter core and an outer quartz layer of 25~$\mu$m with the refractive index smaller by 0.02 with respect to the core, corresponding to the numerical aperture $NA\simeq 0.24$. The volumetric sampling fraction, or the relative volume occupied by the fibres, is 31\%. To minimise the effect of cascade electrons moving along fibres, a 2~mm thick layer of copper is attached to the front of the calorimeter and the whole MGC is vertically tilted by 5 degrees. 

In Fig.~\ref{fig:iv}, the results are shown for the doses induced by bremsstrahlung in fibres of a 30 cm $\times$ 30 cm $\times$ 50~cm copper $QFi$ calorimeter, obtained using {\sc Geant4} and the events generated by {\sc Bremge} as above. The maximum dose of about 7~MGy (700 Mrad) is expected in the central fibres per 100~fb$^{-1}$ of $ep$ collisions. Numerous studies have shown that this dose can be well tolerated by quartz detectors~\cite{Mavromanolakis:2004fr,GORODETZKY1993253}. In addition, as the dose rapidly decreases with distance from the photon beam axis (see Fig.~\ref{fig:iv}, distributions on the right), regular lateral shifts of the MGC position can be envisaged to decrease over time the maximal dose by at least an order of magnitude, and make it more uniform across the fibres.

\section{Performance of the copper $\boldsymbol{QFi}$ calorimeter}

\begin{figure}[t]
\mbox{\includegraphics[width=.245\textwidth]{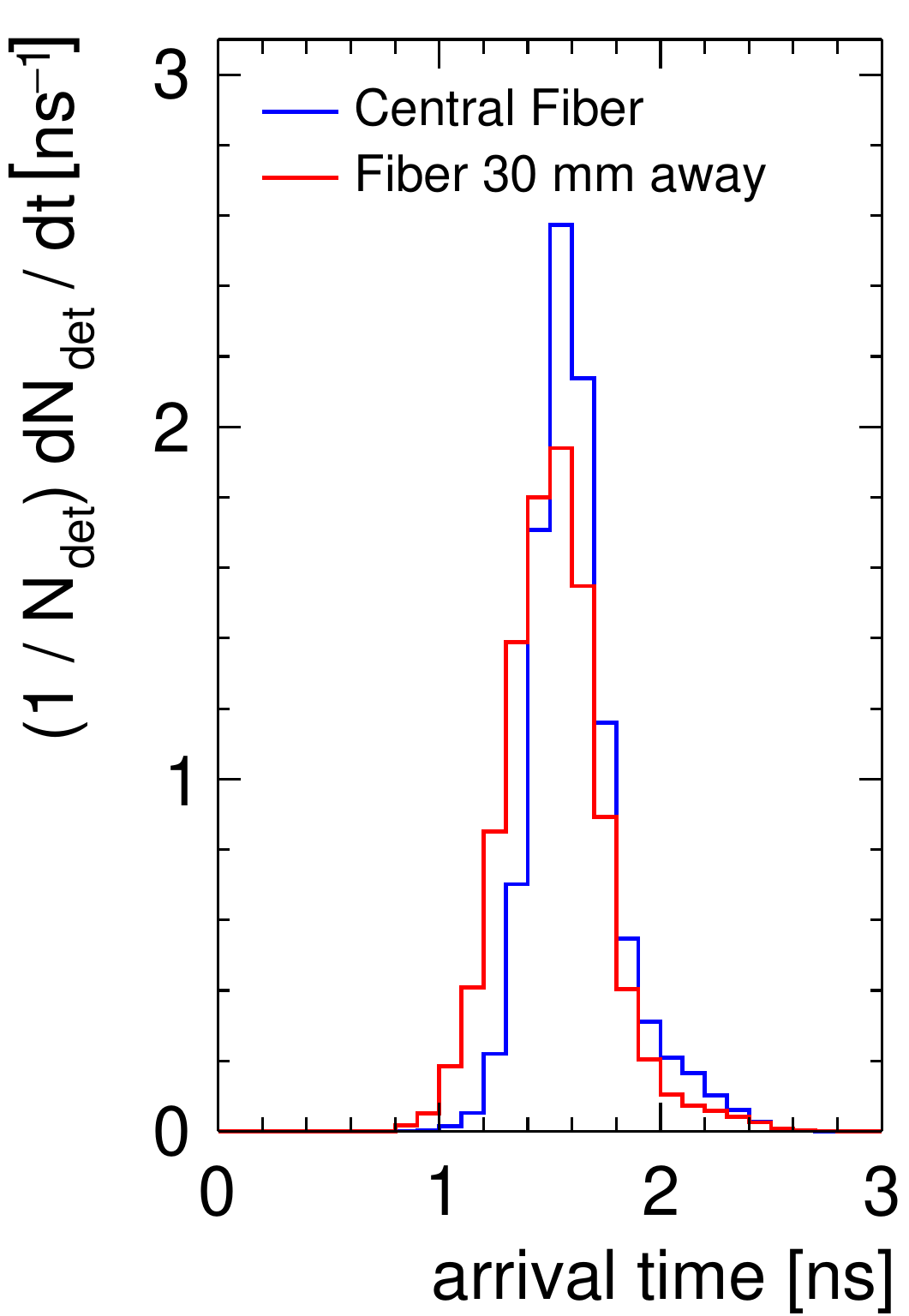}
\includegraphics[width=.73\textwidth]{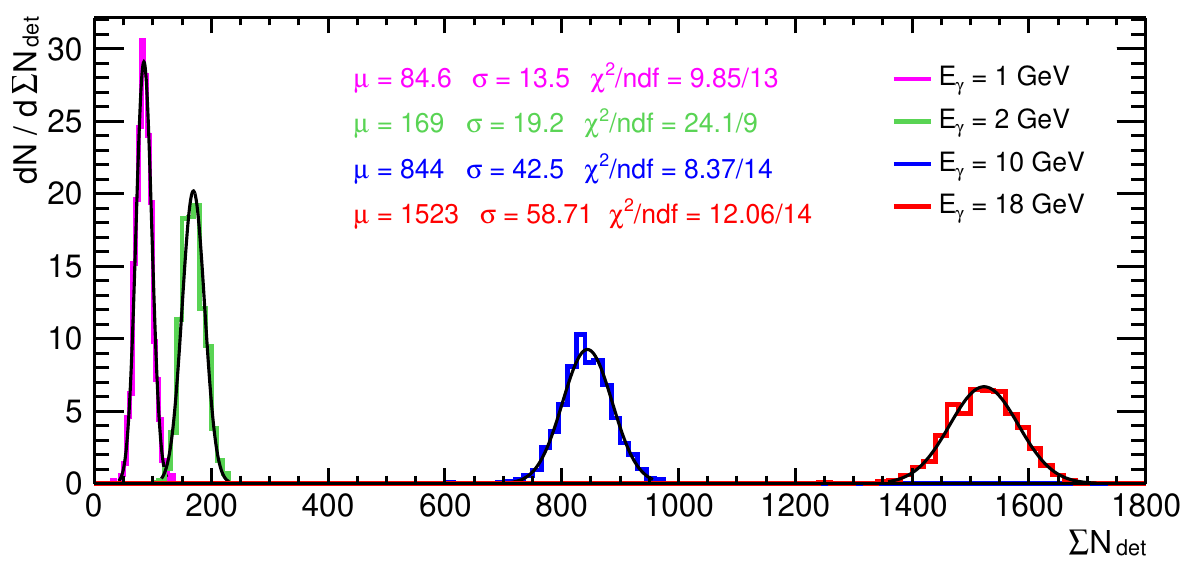}}
\caption{(left) Distributions of Cherenkov photon arrival times to SiPMs. (right) Distributions of the total number of detected photons for 1, 2, 10 and 18~GeV primary photons hitting the $Cu$-quartz ``spaghetti" calorimeter centrally with 1.25~~mm fibres; the calorimeter is tilted vertically by 5 degrees.\label{fig:nPhot}}
\end{figure}

To estimate the performance of the copper $QFi$ calorimeter, special {\sc Geant4} simulations of the electromagnetic cascades, including the generation and propagation of Cherenkov light along the fibres, were performed. The Cherenkov photon wavelengths were limited to the $250<\lambda<650$~nm range, and the number of such photons that exit the end of the fibres was counted. Finally, the number of detected photons was obtained by assuming a Photon Detection Efficiency (PDE) of 0.2, which is a rather conservative estimate of the expected performance of the Silicon Multipliers (SiPMs) proposed as Cherenkov photon sensors. 
\begin{figure}[b]
\centering
\includegraphics[width=.49\textwidth]{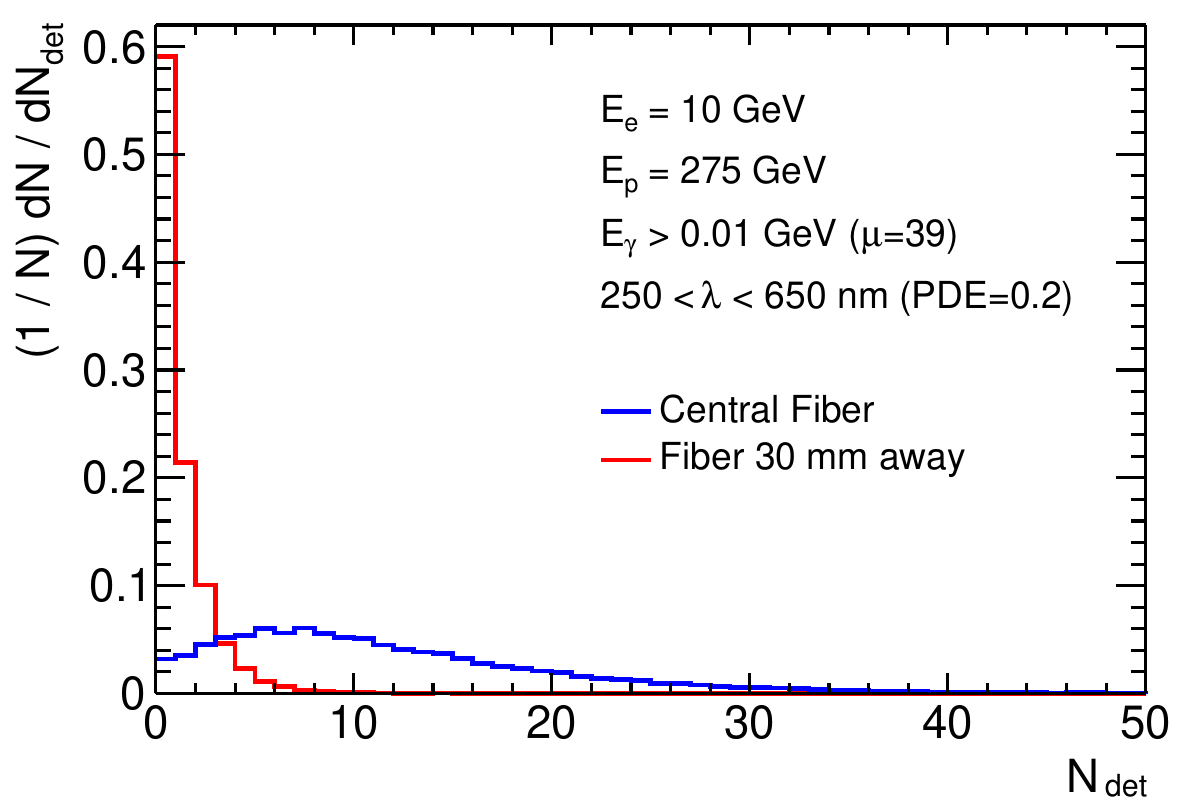}
\hspace{0mm}
\includegraphics[width=.49\textwidth]{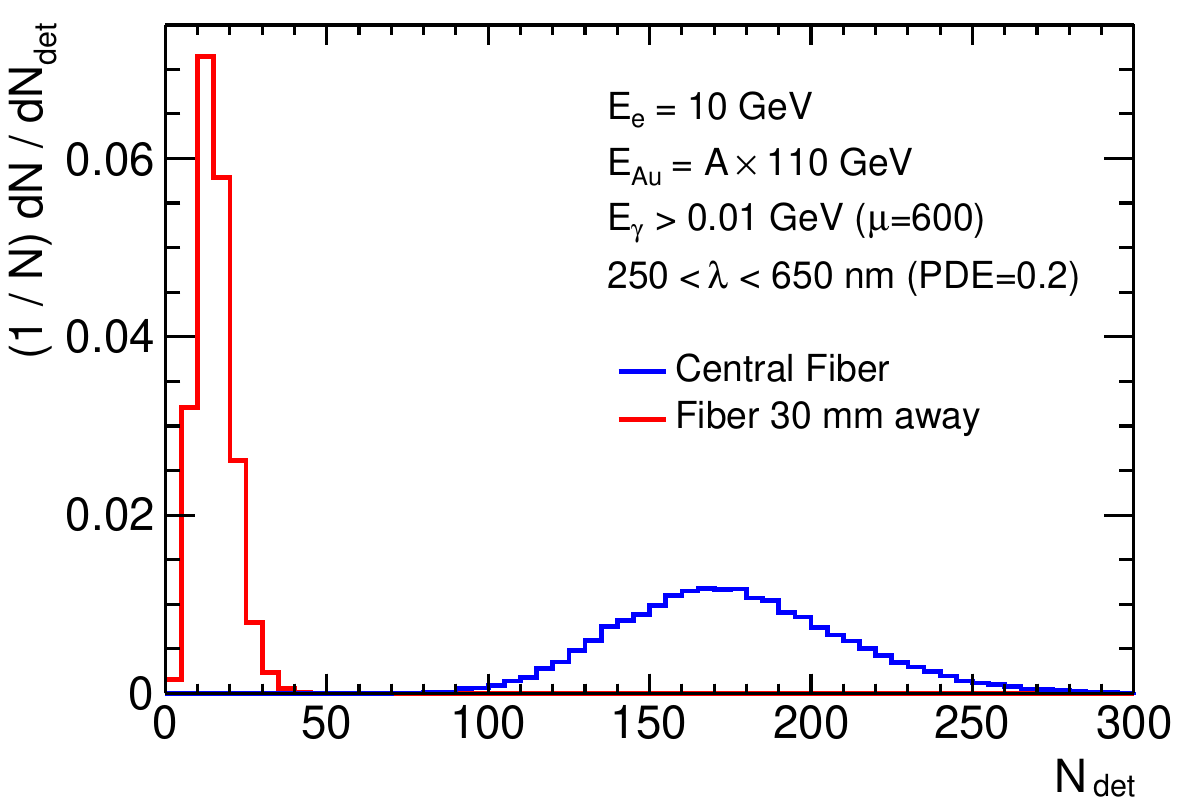}
\caption{Distributions of the number of detected Cherenkov photons exiting central and peripheral quartz fibres for a 10~GeV electron beam at the nominal $ep$ (left) and $e$Au (right) luminosity, assuming PDE=0.2.
\label{fig:v}}
\end{figure}
The results reported in Fig.~\ref{fig:v} show the signal occupancy (i.e. fraction of events with $N_{det}>0$) in the central fibres very close to 100\%, and also for the peripheral ones in the $e$Au case. This, in turn, requires the use of readout electronics capable of measuring the full signal within a 10~ns-wide time window. As shown in Fig.~\ref{fig:nPhot}(left), variations of the Cherenkov photon arrival time to the SiPMs are well contained within that window. It should be noted here that in the final MGC design, numerous fibres will be bundled together and readout by a single~SiPM.

%
\begin{figure}[b]
\centering
\includegraphics[width=.49\textwidth]{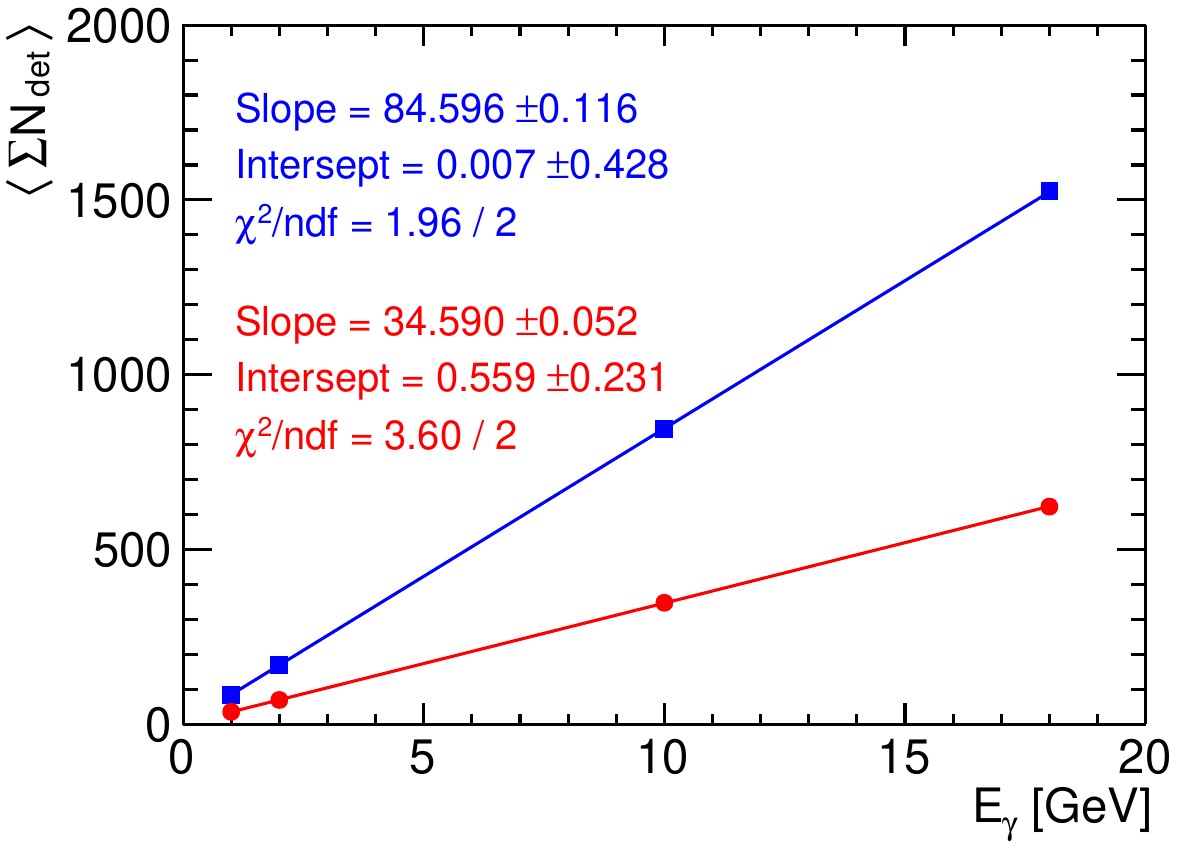}
\hspace{0mm}
\includegraphics[width=.49\textwidth]{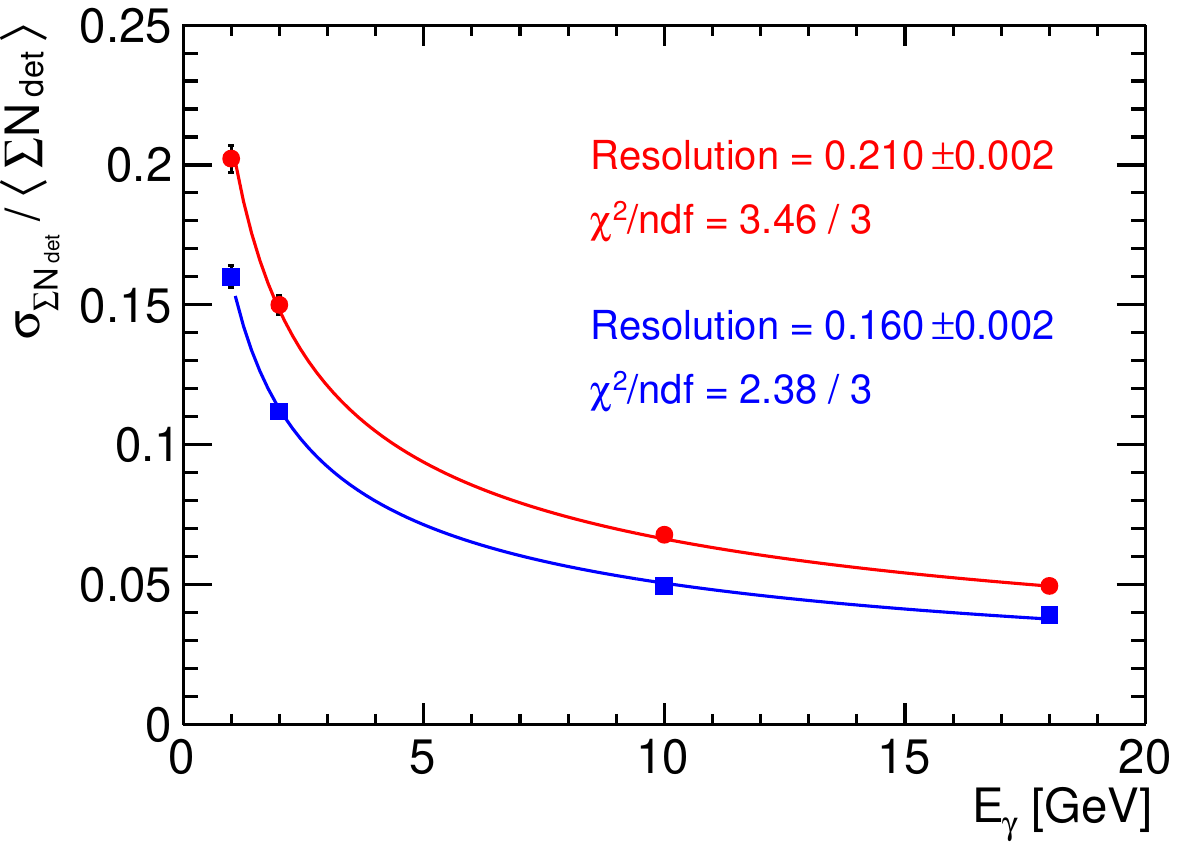}
\caption{(left) The linearity, and (right) the relative energy resolution of a copper $QFi$
``spaghetti" calorimeter as a function of primary photon energy; for the nominal 1.25~mm fibers (blue) and for the case of 0.5~mm diameter fibres with a volumetric sampling fraction of 15\% (red).\label{fig:res}}
\end{figure}

In Fig.~\ref{fig:nPhot}(right), distributions of total numbers of detected Cherenkov photons $\Sigma N_{det}$ are shown for several fixed energies of primary photons. It should be stressed that even for the lowest energy of 1~GeV the $\Sigma N_{det}$ distribution is perfectly Gaussian. 

In Fig.~\ref{fig:res}, the linearity of the MGC response as well as its energy resolution are shown, also for the case of 0.5~mm diameter fibres with a volumetric sampling fraction of only 15\%. The expected linearity of the MGC is excellent, and the relative energy resolution $\sigma_E/E = 0.16/\sqrt{E\mathrm{[GeV]}}$ is sufficient to allow for precise data-driven $in~situ$ re-calibrations of its energy scale.

More studies are needed to optimise the fibre geometry of the MGC and to account for relevant details of the actual setup of luminosity measurement at the EIC, as well as to design the appropriate readout electronics. In that context, the use of flash analog-to-digital converters (FADCs) with a high sampling rate of 200~MHz is foreseen.

\section{Conclusions}
The proposed copper-quartz spaghetti calorimeter is a promising candidate for detecting direct bremsstrahlung photons at the EIC, as indicated by the initial studies reported above. On the one hand, it is expected to withstand enormous event rates, and on the other, still measure the energy of a single photon with a good resolution, allowing for an accurate data-driven calibration, which is essential for a precise determination of the absolute EIC luminosity. In addition, given the extremely high statistics of the measured photons, this will also provide a unique opportunity for a very precise determination of the relative bunch-to-bunch luminosity. The Monte Carlo simulation studies need to be followed up and verified by testing the calorimeter prototype; however, the major remaining challenge is designing appropriate, highly performant SiPM readout electronics.

\section*{Acknowledgements}
Y. Ali, A. Kowalewska, B. Pawlik and K. Piotrzkowski appreciate the financial support of the Polish National Agency for Academic Exchange (NAWA) under grant number BPN/PPO/2021/1/00011.
KP has been supported also by the AGH IDUB grant.
The National Science Centre of Poland partly supported this work under grant number UMO-2024/53/B/ST2/03257.

\appendix
\section{\sc\textbf{Bremge}}
{\sc Bremge} is a Monte Carlo generator of high-energy electron-proton, electron-nucleus and electron-atom(gas) brems\-strah\-lung events. It was initially written in {\sc Fortran77} for studying bremsstrahlung at HERA \cite{Piotrzkowski:1991ym}. A C++ version {\sc BremgeC} has been recently developed \cite{Chwastowski:BremgeC}, which provides output in the {\sc HepMC3} format.

The unweighted events are generated according to the Bethe-Heitler ultra-relativistic differential cross section calculated in the Born approximation and the small-angle approximation. Higher-order diagrams contribute less than 1\% and are therefore not considered. As the energy transfer to the proton or nucleus (i.e the target recoil) is very small, it has been neglected; hence, the sum of photon and scattered electron energies is equal to the electron beam energy. Finally, the cross-section corrections due to beam polarisations~\cite{Gangadharan:2023pmm} and the hadronic structure~\cite{vanderHorst:1990uf,Haas:2010bq}, which are highly suppressed, are also neglected.

\begin{figure}[ht]
\centering
\includegraphics[width=.49\textwidth]{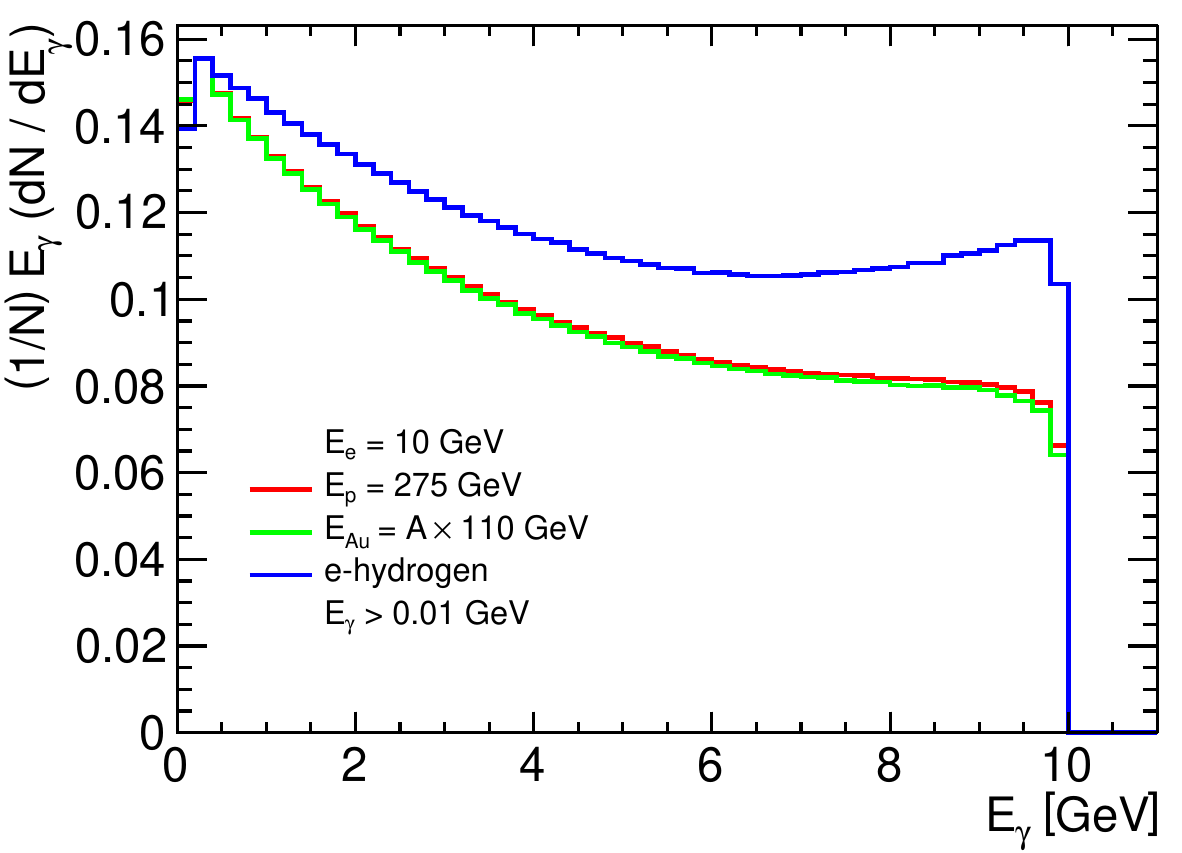}
\hspace{0mm}
\includegraphics[width=.49\textwidth]{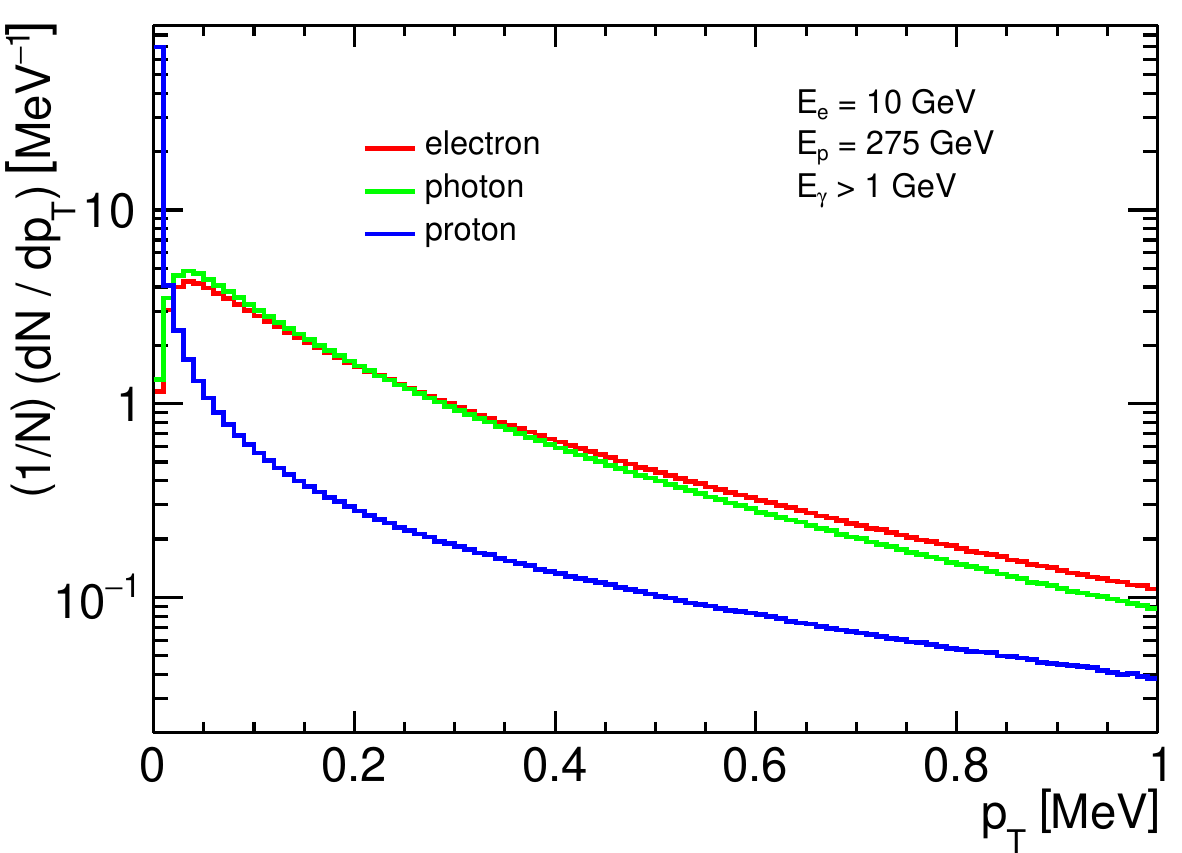}

\caption{Normalized photon spectra $\frac{E_\gamma}{N}\frac{dN}{dE_\gamma}$ (left) for the 10~GeV electron beam $ep$, $e$Au and $e$-hydrogen ($e$-gas) collisions, and $\frac{1}{N}\frac{dN}{dp_T}$  (right) for the radiated photons and the scattered electrons and protons for the $ep$ collisions; no beam-size effect is considered and 100 million events were generated for each sample.\label{fig:a}}
\end{figure}

The user is expected to choose the type of process and provide the mass and charge of the nucleus, the energies of electron and proton/nucleus beams, and the requested range of energies for bremsstrahlung photons. The generator returns the four-momenta of the bremsstrahlung photon and of the scattered electron and proton (or nucleus). 
{\sc Bremge} is very fast, and in Fig.~\ref{fig:a} several generated high-statistics distributions are shown, for example.

\bibliographystyle{JHEP}
\bibliography{biblio.bib}


\end{document}